# An alternative method of adjusting for multiple comparison in medical research

Jiale Li[1*], Zimu Wei[2]

(1[*].Southern Medical University, Magang Avenue, Ronggui Street, Shunde District, Foshan City, Guangdong, P. R. of China; 2. Tianjin Nankai High School, No. 22, Nankai Fourth Road, Nankai District, Tianjin City, P. R. of China. )

**Abstract**

**Background**

Most methods of adjusting for multiplicity focus primarily on controlling type I errors and rarely consider type II errors. We propose a new method that considers controlling for false-positive findings while ensuring sufficient statistical power.

**Methods**

We proposed a new method for multiple corrections called (Beta-exponential Adjustment, BEA) that considered the statistical power to control for type I errors while also considering the probability of type II errors. We conducted simulation studies to evaluate the performance characteristic of multiple testing correction procedures. We calculated sensitivity, specificity, and power separately for different sample sizes and number of biomarkers and compared them with the Bonferroni, Holm, and Benjamini-Hochberg (BH) correction methods.

**Results**

The results demonstrated that our proposed BEA correction method exhibited the highest sensitivity at different sample sizes and biomarkers (e.g., sensitivity: BEA 0.8 versus BH

0.62 at sample size at 1000, tested biomarkers at 1000 and positive rate at 30%). With different sample sizes and number of biomarkers, the BEA correction method demonstrated comparable specificity compared with traditional methods. Moreover, we observed that the BEA-corrected had the highest statistical power than other methods, when the outcome was relatively rare.

**Conclusion**

We proposed the BEA multiple correction method to adjust for multiple comparisons while considering statistical power. The BEA method demonstrated a higher sensitivity, comparable specificity, and higher statistical power, compared with traditional correction methods in different conditions. The BEA correction method can be an alternative of traditional methods of adjusting for multiplicity, especially in studies with small sample size, rare outcomes, or substantial number of biomarkers.

# INTRODUCTION

P-value is an important concept in testing hypothesis and is the probability of observing the current sample data given that the original hypothesis ($H_0$) is true.[1,2] In medical research, p-value is an important parameter to draw a conclusion (i.e., accept or reject a hypothesis). In statistical tests, the level of significance (α) is determined first and then compared to the p-value to conclude whether there is statistical significance.[3] In scientific research, several hypotheses often need to be tested in a single study. When performing multiple tests, making judgement based only on the significance level of a single test greatly increases the risk of type I errors (false positive).[4] Multiple comparison corrections are usually recommended to avoid rejecting the true $H_0$.[5] However, the trade-offs of benefit versus detriment of multiple adjustments has been controversial. For instance, Rothman K spoke against adjustment of multiplicity since reduction in type I errors may be accompanied by increase of type II errors, and the fear of type I errors should not obscure the potential for discovering novelty.[6] Such argument encouraged a more balanced approach to scientific research, taking into account both type I and type II errors, rather than being overly biased in favor of avoiding one type of error.

In medical research, multiple comparisons widely existed in various areas of experimental or observational data, such as genomics[7], pharmacology[8], epidemiology[9], as well as imaging[10]. A variety of multiple adjustment methods had already been available[4,5,11]. The Bonferroni correction is a classical correction for multiple comparisons by setting the significance level of each test to $\frac{\alpha}{m}$, where α is the original significance level (e.g., 0.05) and m is the number of independent hypothesis tests[12,13]. The Bonferroni correction provides a tight control of the overall type I error rate, but is conservative and may lead to an increase in the type II error rate and loss of some essential findings[14]. The Holm-Bonferroni method is a modification of the Bonferroni correction designed to control the overall error rate while reducing its conservatism[15]. All p-values are sorted from smallest to largest and then corrected stepwise starting with the smallest p-value. For the kth p-value, the corrected significance level is $\frac{\alpha}{m-k+1}$. If the p-value is less than the corrected significance level, the original hypothesis is rejected. In addition, the Benjamini-Hochberg method (BH method) can be applied, which can control the False Discovery Rate (FDR) [16]. Sorting the p-values from smallest to largest, the corrected p-value = $p_k \times \frac{m}{k}$. It is relatively efficient and allows for a certain percentage of false positives, which makes it particularly suitable for large-scale data analysis.

Although various methods had been widely used for adjusting multiplicity in medical research, most of them focus on controlling for type I error, with little consideration for type II error, which may result in under-powered estimates, particularly in studies with rare outcomes. To address the knowledge gap, we developed a novel method that considered for controlling for false positive discovery while ensuring sufficient statistical power.

## METHOD

### Formula

Assuming that there are $m_0$ independent hypothesis tests, n of them reached unadjusted significance level α (i.e., α=0.05; p<0.05 indicates statistical significance), $b_1$ is the proportion of number of the positive findings (unadjusted significance level) to number of

all tests:

$$b_1 = \frac{n}{m_0} \quad (1)$$

We then proposed the following theory that number of adjustment ($m_1$) for multiple comparisons should be dependent on the positive findings:

$$m_1 = m_0 \times b_1 = m_0 \times \frac{n}{m_0} \quad (2)$$

When all tests reached unadjusted significance level (i.e., $m_0 = n$, $b_1=1$), we need to consider adjustment for all tests (i.e., $m_1 = m_0$, similar as Bonferroni correction); When no test reached unadjusted significance level (i.e., $n=0$, $b_1=0$), we do not need to adjust for any test ($m_1=0$, similar as Rothman theory of no adjustment).

This above formula implied that the number of corrections depends on the number of positive results. Next, we additionally incorporate statistical power and $\beta$ (type II error) into the above formula;

$$M_2 = m_0 \times (\frac{n}{m_0})^x \quad (3)$$

in which

$$x = \frac{1}{1-\beta}$$

The formula took statistical power into account, controlling for type I errors while also considering the probability of type II errors. The number of tests decrease with the increasing of type II error (i.e., $\beta$). As statistical power=1-$\beta$, X=1/statistical power. If the $\beta$ is close to 0, then $M_2$ is close to n; if the $\beta$ is close to 1, then X is close to ∞ and $M_2$ is close to 0 ($\frac{n}{m_0} \leq 1$).

**Simulation data**

We conducted simulation studies to evaluate the performance characteristic of multiple testing correction procedures. The simulated data set includes the following information: (1) binary disease status for patients (Number=n) and (2) expression levels of biomarkers (Number=m) for each patient. For each simulated dataset, we performed logistic regression analyses to quantify biomarker-disease associations, subsequently obtaining original p-values for each biomarker test. Five distinct multiple testing correction procedures were applied: Bonferroni, Holm, Benjamini-Hochberg, and our novel proposed method (Beta-exponential Adjustment, BEA), with the initial level of significance was set at 0.05. In the formula of our new method, the β-value for calculating the sample size was set to 0.8. We conducted multiple corrections on simulated data with different sample sizes (n) and numbers of biomarkers (m), and compared the sensitivity, specificity, and power of each correction method. To enhance the credibility of the results, we simulated each data set with the same number of patients (i.e., n) and the same number of biomarkers (i.e., m) for 100 times. We also explored the impact under the condition with p-values less than 0.05 on the total number of tests on sensitivity, specificity, and power. We use the base package *stats* in R for previous multiple correction methods.

**Sensitivity and specificity**

After regressing the simulated data, we randomly labeled the tests with original p-values < 0.05 to determine whether they were true positives or true negatives, and we considered all tests with original p-values > 0.05 as true negatives. Subsequently, we applied multiple correction methods to adjust the level of significance and then calculated the sensitivity and specificity for each method based on the corrected results. Sensitivity is defined as the true positive rate, while specificity is the true negative rate.

**Power**

We substituted the adjusted the level of significance (α_corrected) of each method into the formula for calculating sample size and obtained a new beta value (β_corrected), where power = 1-β_corrected. The original beta in the sample size calculation formula was set to 0.8.

Statistical analyses were all performed using RStudio (version 4.4.1).

**RESULTS**

**Sensitivity comparison**

Firstly, we compared the sensitivity using Bonferroni, Holm, BH and BEA corrections in simulated cohort with different sample sizes and number of biomarkers. The results demonstrated that our BEA correction method exhibited the highest sensitivity at different sample sizes and biomarkers (Figure 1).

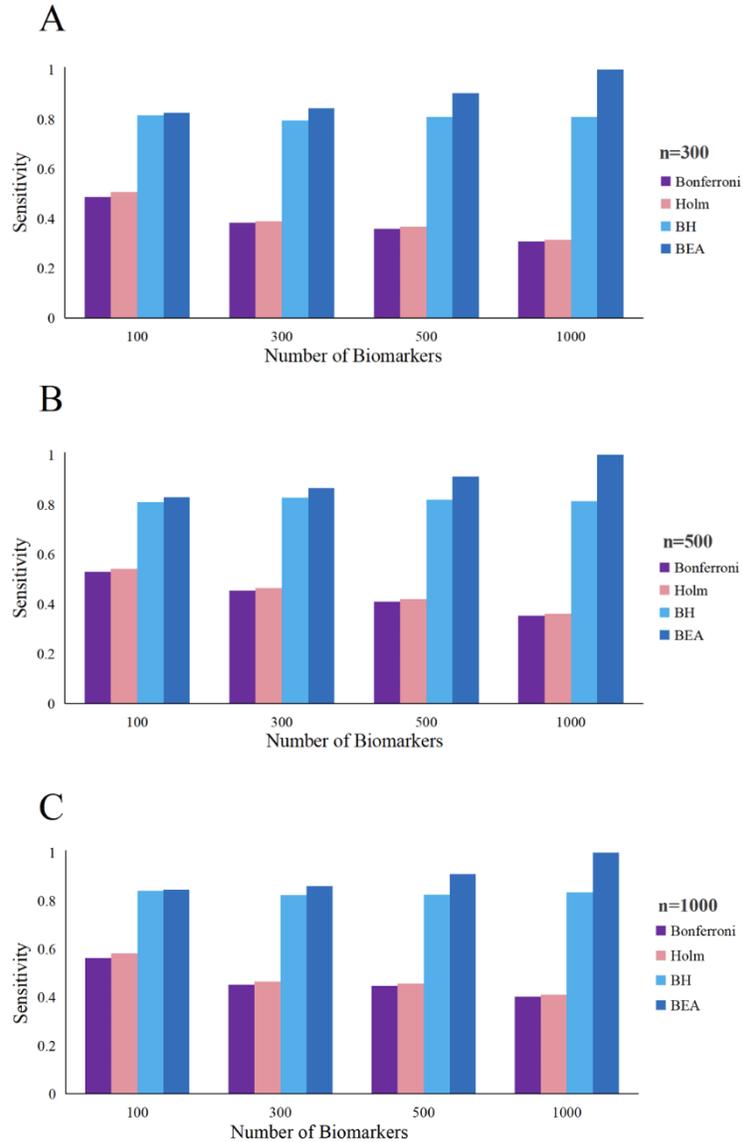

Figure 1. **Sensitivity comparison among the Bonferroni, Holm, BH and BEA corrections under different sample size, number of Biomarkers, with 30% of original p-values less than 0.05 (single-armed queue).**

With the addition of biomarkers, the sensitivity of the BEA-correct ed method gradually increased and showed increasing advantages over the other three methods. With a sample size of 1,000 and a positive rate of 30%, along with biomarkers at 100, 300, 500, and 1,000 respectively, the sensitivities of our BEA correction method were 0.85, 0.86, 0.91, and 1.00, respectively, showing a significant up trend. In contrast, the sensitivities of the BH method were 0.84, 0.82, 0.83, and 0.84 under the same conditions, respectively, and their values did not fluctuate significantly with the increase of biomarkers. The sensitivity of the Holm and Bonferroni is lower, falling below 60%. Notably, our BEA method has a sensitivity of 1.00 when the number of biomarkers is substantial (e.g., 1,000), suggesting its advantages in high-throughput data analysis. When the sample size is 300 or 500, the sensitivity differences of various correction methods are similar to those observed by the sample size is 1000. In cohorts with different positive findings (i.e., different proportions of raw p-values less than 0.05, we set different positive rate at 30%, 40%, 60% and 70%, respectively) (Figure 2), the sensitivity of the corrected BEA was similar as the BH correction method, but was higher than the Bonferroni and Holm corrections. With a sample size of 1,000 and biomarker thresholds set at 1,000, the positive rates were established at 30%, 40%, 60%, and 70%, respectively. The sensitivities for BEA were 1.00, 0.94, 0.76, and 0.76, respectively, while those for BH were 0.84, 0.92, 0.96, and 0.98, respectively. The difference suggests that the BEA method is more suitable for scenarios with low positive rates (e.g., rare disease), while the BH method performs better in research setting with high positive rate. The sensitivity of both Holm and Bonferroni methods was found to be lower than those of BEA and BH.

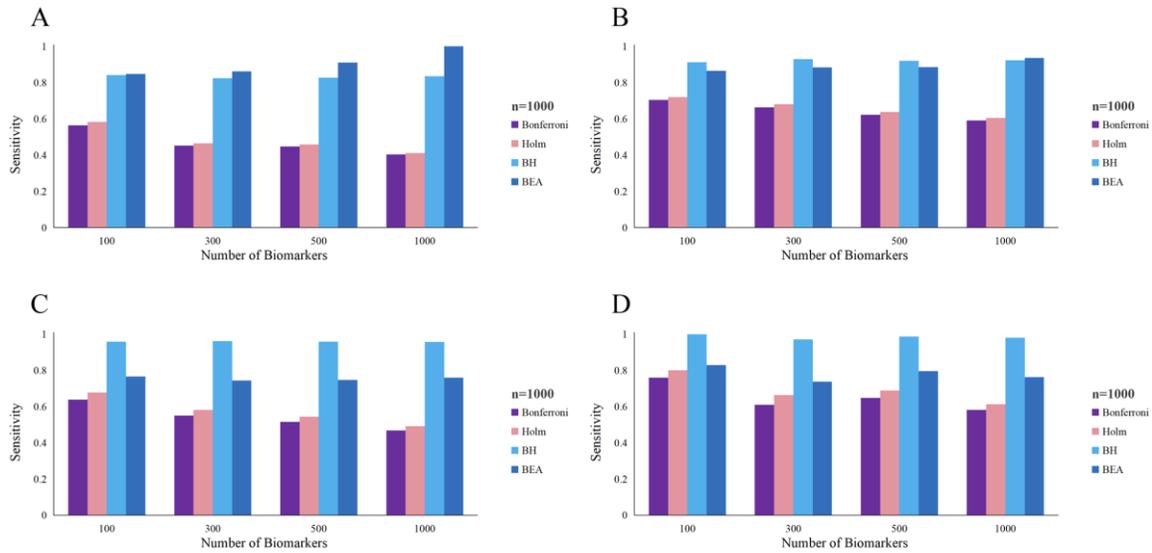

Figure 2. **Sensitivity comparison among the Bonferroni, Holm, BH and BEA corrections under different number of Biomarkers and percentage of original p-values less than 0.05, with the sample sizes of 1000.** (A) 30% of original p-values less than 0.05, (B) 40% of original p-values less than 0.05, (C) 50% of original p-values less than 0.05, (D) 60% of original p-values less than 0.05.

**Specificity comparison**

We further compared the specificity of Bonferroni, Holm, BH, and BEA corrections with different sample sizes and different number of biomarkers. Under different sample sizes different number of biomarkers and different positive rate, the specificity was comparable among the studied methods. (Figure 3-4).

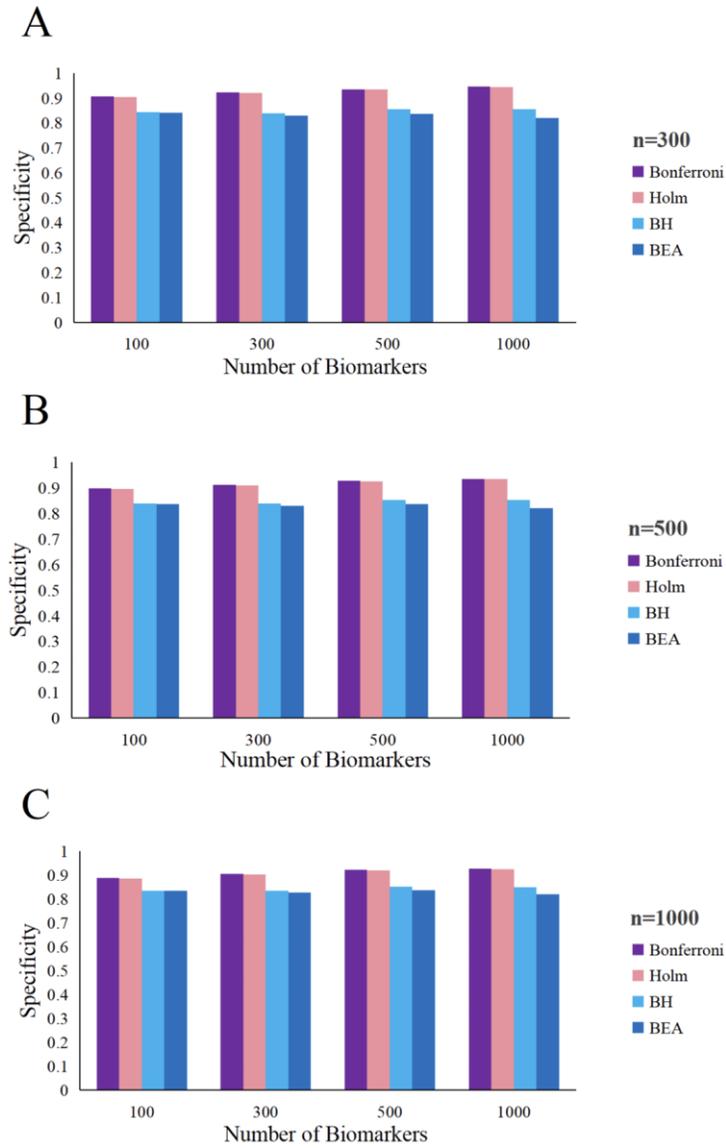

Figure 3. **Specificity comparison among the Bonferroni, Holm, BH and BEA corrections under different sample size, number of Biomarkers, with 30% of original p-values less than 0.05.**

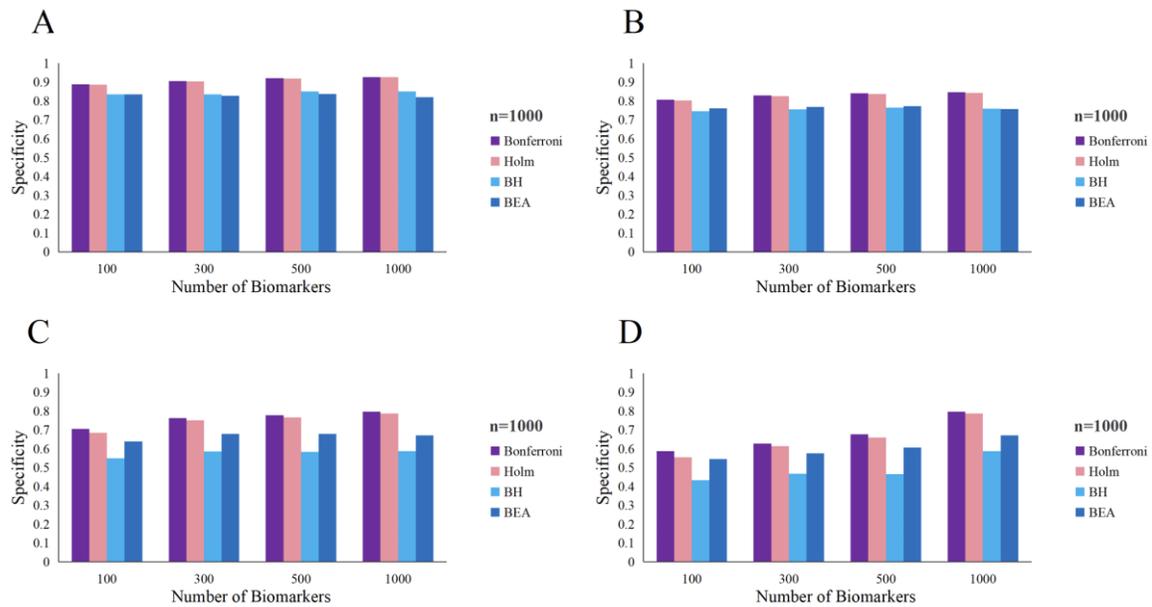

Figure 4. **Specificity comparison among the Bonferroni, Holm, BH and BEA corrections under different number of Biomarkers and percentage of original p-values less than 0.05, with the sample size of 1000.** (A) 30% of original p-values less than 0.05, (B) 40% of original p-values less than 0.05, (C) 50% of original p-values less than 0.05, (D) 60% of original p-values less than 0.05.

**Power comparison**

Finally, we compared the power of Bonferroni, Holm, BH, and BEA corrections. We observed that the BEA method had the highest power than other methods under number of biomarkers, especially for a large number of biomarkers (Figure 5).

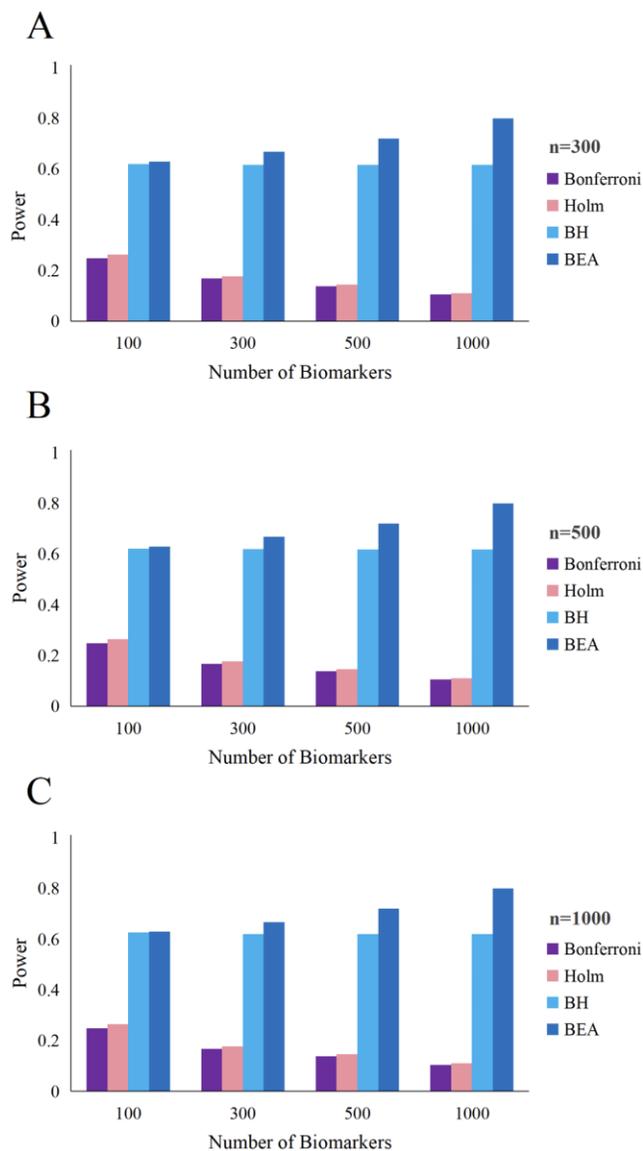

Figure 5. **Power comparison among the Bonferroni, Holm, BH and BEA corrections under different sample size, number of Biomarkers, with 30% of original p-values less than 0.05.**

As the number of biomarkers grew, the power of the BEA correction method increased, and its advantages became apparent. With a sample size of 1000, the positive rate set as 30%, biomarkers at 100, 300, 500, 1000, respectively, the power of BEA are 0.63, 0.67, 0.72, and 0.80, respectively, while that of BH are 0.63, 0.62, 0.62, and 0.62, respectively.

The power of both Holm and Bonferroni is lower than that of these two methods, even less than 30%. Provided that the sample size and the quantity of biomarkers are kept identical, modifying the positive rate causes the power of the BEA and BH to display a comparable trend to that seen with alterations in sensitivity. With a sample size of 1,000 and biomarkers set at 1,000, the positive rates were established at 30%, 40%, 60%, and 70%, respectively. The power for BEA were 0.80, 0.70, 0.42, and 0.31, respectively, while those for BH were, respectively, 0.62, 0.67, 0.73, and 0.75. It is worth noting that when the original p-value positive rate is increased to 70%, although the efficacy of the BEA method (0.31) is lower than that of the BH method (0.75), its performance under the condition of low positive rate (e.g., 0.80 at 30%) is still significantly better than that of the traditional method, indicating that the BEA method is more suitable for the identification of sparse signals in exploratory studies.

**DISCUSSION**

In this study we proposed a novel method (BEA) for multiple corrections that considered both the level of significance and statistical power. Our BEA method demonstrated a higher sensitivity, comparable specificity, and higher statistical power than traditional multiple correction methods, particularly under the condition of small sample size and substantial number of biomarkers.

Multiple test corrections assist to avoid unwarranted positive findings in the analysis[17]. Suitable multiple correction method should be selected in different clinical situation of medical research according to the research purpose, data type, sample size and other factors.

Given the stringency of the Bonferroni correction, it is appropriate to consider when the cost of false positives in a study is high, such as in the diagnosis of certain serious diseases, as any false positives may lead to unnecessary treatment and unaffordable consequences[18]. The Holm correction is more flexible than the Bonferroni correction and can improve test efficacy while controlling the false positive rate[19]. The Benjamini-Hochberg test has high efficacy and allows for a certain percentage of false positives, making the BH correction an appropriate choice when meaningful effects need to be detected in a large number of tests, such as genomics studies[20]. The BEA correction method we proposed is particularly suitable for rare diseases with small samples and enormous number of comparisons.

In addition to the widely used multiplicity adjustment methods described above; some new advanced multiplicity adjustment methods have been proposed. Fixed-order procedure is hypothesis test conducted in a predefined order. Only when the previous hypothesis test rejects the original hypothesis will it proceed to the next hypothesis test, until a certain hypothesis test does not reject the original hypothesis, and the final inferential conclusion is that the significance conclusions of the previous hypothesis tests are all accepted[21, 22]. One of its obvious drawbacks is that it does not allow further testing once the hypothesis has not been rejected. Wiens introduced a fallback procedure to address this major shortcoming by allowing all hypotheses to be tested in a pre-specified sequence even if the initial hypotheses are not rejected[23]. In addition, Li and Mehrotra introduced an adaptive alpha allocation approach to allow the significance level of later tested hypotheses in the series to depend on the level of evidence for testing earlier hypotheses[24]. The adaptive alpha allocation approach strongly controlled the familywise error rate (FWER) for two or more independent endpoints. However, the structure of the above calibration methods was too complex relative to our proposed BEA method and did not give much consideration to evaluate validity.

The most obvious advantages of our BEA methods are the combination of statistical power and control for false positive finding, which is especially suitable for medical research with small sample size, rare outcomes, or substantial number of tests. However, our methods also had limitations. One limitation is that increase statistical power is inevitably at the cost of relatively higher rate of false positive findings. In addition, another limitation is that the process of our BEA method is more complicated than Bonferroni, Holm, and BH methods.

In conclusion, we proposed the BEA multiple correction method to adjust for multiple comparisons while considering statistical power. The BEA method demonstrated a higher sensitivity, comparable specificity, and higher statistical power, compared with traditional correction methods in different conditions, The BEA correction method can be an alternative of traditional methods of adjusting for multiplicity, especially in studies with small sample size, rare outcomes, or substantial number of biomarkers.

**Declarations**

**Author contributions:** Zimu Wei conceived the idea, performed the analyses, and draw the figures, and drafted the manuscript. Jiale Li revised the manuscript and draw the figures

**Data and code sharing statement:** The data used in this paper are randomly generated simulations. The original code in this paper is available by contacting the corresponding author.

**Declaration of interests:** The authors declare no conflict of interest.

**Ethics:** Not applicable.

**REFERENCE**

1. Thiese MS, Ronna B, Ott UJJoTD. P value interpretations and considerations. J Thorac Dis. 2016;8(9):E928-E31.

2. Boos DD, Stefanski LA. P-Value Precision and Reproducibility. The American Statistician. 2011;65(4):213-21.

3. Greenland S, Senn SJ, Rothman KJ, Carlin JB, Poole C, Goodman SN, et al. Statistical tests, P values, confidence intervals, and power: a guide to misinterpretations. European Journal of Epidemiology. 2016;31(4):337-50.

4. Bender R, Lange S. Adjusting for multiple testing--when and how? J Clin Epidemiol. 2001;54(4):343-9.

5. Gordi T, Khamis H. Simple solution to a common statistical problem: interpreting multiple tests. Clin Ther. 2004;26(5):780-6.

6. Rothman KJ. No adjustments are needed for multiple comparisons. Epidemiology. 1990;1(1):43-6.

7. Sham PC, Purcell SM. Statistical power and significance testing in large-scale genetic studies. Nature Reviews Genetics. 2014;15(5):335-46.

8. Petschner P, Bagdy G, Tóthfalusi L. The problem of small "n" and big "P" in neuropsycho-pharmacology, or h ow to keep the rate of false discoveries under control. Neuropsychopharmacol Hung. 2015;17(1):23-30.

9. Ottenbacher KJ. Quantitative evaluation of multiplicity in epidemiology and public hea lth research. Am J Epidemiol. 1998;147(7):615-9.

10. Alberton BAV, Nichols TE, Gamba HR, Winkler AM. Multiple testing correction

over contrasts for brain imaging. Neuroimage. 2020;216:116760.

11. Yoav B, Daniel Y. The control of the false discovery rate in multiple testing under dependency. The Annals of Statistics. 2001;29(4):1165-88.

12. Curtin F, Schulz P. Multiple correlations and Bonferroni's correction. Biological Psychiatry. 1998;44(8):775-7.

13. Sedgwick P. Multiple hypothesis testing and Bonferroni's correction. BMJ. 2014;349:g6284.

14. Perneger TV. What's wrong with Bonferroni adjustments. BMJ. 1998;316(7139):1236.

15. Holm S. A Simple Sequentially Rejective Multiple Test Procedure. Scand J Stat 1979;6:65-70.

16. Ghosh D. Incorporating the Empirical Null Hypothesis into the Benjamini-Hochberg Procedure. Statistical Applications in Genetics and Molecular Biology. 2012;11(4).

17. Menyhart O, Weltz B, Győrffy B. MultipleTesting.com: A tool for life science researchers for multiple hypothesis testing correction. PLOS ONE. 2021;16(6):e0245824.

18. Armstrong RA. When to use the Bonferroni correction. Ophthalmic and Physiological Optics. 2014;34(5):502-8.

19. Fu G, Saunders G, Stevens J. Holm multiple correction for large-scale gene-shape association mapping. BMC Genetics. 2014;15(1):S5.

20. David SR, James MSW, Aaditya R. Online Multiple Hypothesis Testing. Statistical

Science. 2023;38(4):557-75.

21. Westfall PH, Krishen A. Optimally weighted, fixed sequence and gatekeeper multiple testing procedures. Journal of Statistical Planning and Inference. 2001;99(1):25-40.

22. Alosh M, Bretz F, Huque M. Advanced multiplicity adjustment methods in clinical trials. Statistics in Medicine. 2014;33(4):693-713.

23. Wiens BL. A fixed sequence Bonferroni procedure for testing multiple endpoints. Pharmaceutical Statistics. 2003;2(3):211-5.

24. Li J, Mehrotra DV. An efficient method for accommodating potentially underpowered primary endpoints. Statistics in Medicine. 2008;27(26):5377-91.